\documentclass[pra,aps,twocolumn,showpacs]{revtex4}
\usepackage{amsmath,amssymb,bm}
\usepackage{epsfig}
\usepackage{graphicx}
\usepackage{psfrag}

\def\unity{\mbox{\small 1} \!\! \mbox{1}}

\begin{document}
\title{Creating large noon states with imperfect phase control}  

\author{Pieter Kok} \email{pieter.kok@materials.ox.ac.uk}
\affiliation{Department of Materials, Oxford University, Parks Road, Oxford
  OX1 3PH, United Kingdom.}  


\begin{abstract}
 \noindent 
 Optical noon states $(|N,0\rangle + |0,N\rangle)/ \sqrt{2}$ are an
 important resource for Heisenberg-limited metrology and quantum
 lithography. The only known methods for creating noon states with 
 arbitrary $N$ via linear optics and projective measurments seem to
 have a limited range of application due to imperfect phase control.
 Here, we show that bootstrapping techniques can be used to create
 high-fidelity noon states of arbitrary size. 
\end{abstract}

\date{\today}
\maketitle

\noindent 
\paragraph*{Introduction}
An important part of quantum information processing is quantum
metrology and quantum lithography. We speak of quantum---or
Heisenberg-limited---metrology when systems in quintessentially
quantum mechanical states are used to reduce the uncertainty in a
phase measurement below the shot-noise limit. If $\phi$ is the phase
to be estimated, and $N$ is the number of independent trials in the
estimation, the shot-noise limit is given by $\Delta\phi =
1/\sqrt{N}$. In quantum mechanics, the $N$ trials can be correlated
such that the limit is reduced to $\Delta\phi = 1/N$
\cite{caves81,yurke86}. It is believed that this is the best phase
sensitivity achievable in quantum mechanics. 

In optics, $\phi$ may represent the length change in the arm of an
interferometer searching for gravity waves. When coherent
(laser) light is used, the phase sensitivity is $1/\sqrt{\bar{n}}$,
where $\bar{n}$ is the average number of photons in the beam. If, on
the other hand, special quantum states of light are used, the phase
sensitivity can be improved. One of such states is the so-called {\em
  noon} state:
\begin{equation}\label{noon}
 |N::0\rangle_{ab} \equiv \frac{1}{\sqrt{2}} \left( |N,0\rangle_{ab} +
   |0,N\rangle_{ab} \right) . 
\end{equation}
If one of the modes experiences a phase shift $\phi$, the state becomes
$(|N,0\rangle + e^{iN\phi}|0,N\rangle)/\sqrt{2}$. The enhanced phase
leads to an increased phase sensitivity of $\Delta\phi = 1/N$
\cite{bollinger96}, which can easily be verified by noting that a
phase shift of $\pi/N$ transforms Eq.~(\ref{noon}) into an orthogonal
state. This means there exists a single-shot experiment that determines
the presence or absence of the phase shift.

Another application that requires the ability to create noon states is
quantum lithography \cite{boto00}. Classical light can write and resolve
features only with size larger than about a quarter of the
wavelength: $\Delta x = \lambda/4$. This is why classical
optical lithography is struggling to reach the atomic level. With the
use of noon states, however, the minimum resolvable feature size
becomes $\Delta x = \lambda/4N$. The same phase enhancement $N\phi$
that gives rise to the Heisenberg limit also enables an unbounded
increase in optical resolution \cite{kok04}. Consequently, noon
states have attracted quite some attention in recent years
\cite{hofmann04,combes05,pezze05,sun05}.

Currently, there are two main procedures for creating noon states:
Kerr nonlinearities \cite{gerry00}, and linear optics with projective
measurements \cite{fiurasek02,lee02,kok02}. Kerr, or optical $\chi^{(3)}$
nonlinearities may in principle yield perfect noon states, but the
small natural coupling of $\chi^{(3)}$ and the unavoidable additional
transformation channels pose a formidable challenge to any practical
implementation. Electromagnetically induced transparencies may be used
to solve this problem \cite{schmidt96}, but even here the creation of
noon states needs nonlinearities with appreciably greater strength
than what has been demonstrated so far.

All methods for creating large noon states with linear
optics and projective measurements use the Fundamental Theorem of
Algebra, which states that every polynomial has a factorization
(see e.g., Ref.~\cite{friedberg89}). In particular, the polynomial
function of the creation operators that generate a noon state is
factorized by the $N$-th roots of unity: 
\begin{equation}\label{eq:factors} \hat{a}^{\dagger N} -
\hat{b}^{\dagger N} = \prod_{k=1}^N (\hat{a}^{\dagger} + e^{2\pi i
(k-1)/N}\, \hat{b}^{\dagger})\; .
\end{equation} Every factor can be implemented probabilistically using
beam splitters, phase shifters, and photo-detection
\cite{fiurasek02,kok02}. Three- and 
four-photon noon states have been demonstrated experimentally by
Mitchell et al.\ \cite{mitchell04} and Walther et al.\
\cite{walther04}, respectively. In this note, I identify a fundamental
problem with the noon-state preparation procedure using linear optics
and projective measurements. In addition, I propose a method that can
be used to circumvent this problem.

\paragraph*{Noisy state preparation}
In practice the phase factor $2\pi(k-1)/N$ in
Eq.~(\ref{eq:factors}) cannot be created with infinite precision. The
accuracy of adjusting the phase is bounded by the limits of
metrology. In order to create noon states, we must be able to tune the
phase shift such that $2\pi (k-1)/N$ and $2\pi k/N$ are well
separated. We thus require the phase error to be smaller than
$2\pi/N$. This is the Heisenberg limit. If our objective is to create
noon states in order to attain the Heisenberg limit, then we encounter
a circular argument. This naive line of reasoning therefore suggests
that the Heisenberg limit cannot be attained this way. In this note,
I quantify the maximum sensitivity using noisy noon states, and
explore a possible way to create high-fidelity noon states of
arbitrary size.

To estimate the effect of imperfect control over the phase
shifts in the preparation process, consider the following
noise model. Every phase in every factor of Eq.~(\ref{eq:factors}) has
a Gaussian distribution with variance $\sqrt{\delta}\,$:
\begin{eqnarray}
 \rho &=& \frac{1}{2N!} \prod_{k=1}^{N}
\int\frac{d\varphi_k}{\sqrt{2\pi\delta}}\, e^{-(\varphi_k - 2\pi
k/N)^2/2\delta} \cr && \qquad \times \left( \hat{a}^{\dagger} +
e^{i\varphi_k}\, \hat{b}^{\dagger} \right) |0\rangle_{ab}\langle 0|
\left( \hat{a} + e^{-i\varphi_k}\, \hat{b} \right) .
\end{eqnarray} 
The variance $\sqrt{\delta}$ is considered
sufficiently small such that the integration can be taken over the
interval $(-\infty,+\infty)$.

To derive the uncertainty in the phase, we adopt the
following measurement model: By virtue of quantum lithography
\cite{boto00}, the noon state can be focussed onto a small region with
width $\pi/N$. In this region, a detector measures the observable
\begin{equation}
 \hat{\Sigma} = |N,0\rangle_{ab}\langle 0,N| + |0,N\rangle_{ab}\langle N,0|\; .
\end{equation} 
For a physical model of such a measurement, see Boto {\em et al}.\
\cite{boto00}. In terms of projection operators, this measurement can
be written as 
\begin{equation}
 \hat{E}_{\pm} = \frac{1}{2} \left( |N,0\rangle \pm |0,N\rangle
 \right)  \left( \langle N,0| \pm \langle 0,N| \right)\; ,
\end{equation}
and the evolution due to the phase shift yields
\begin{equation}
 \rho (\phi) = \left( \unity\otimes e^{i\hat{n}_b\phi}\right)\, \rho\,
 \left( \unity\otimes e^{-i\hat{n}_b\phi} \right), 
\end{equation}
where $\hat{n}_b = \hat{b}^{\dagger}\hat{b}$. The conditional
probability of finding outcome $j$ in a measurement given a phase shift
$\phi$ is then calculated as follows:
\begin{equation}
 p(j|\phi) = {\rm tr}\left[ \hat{E}_j \rho(\phi) \right].
\end{equation}
The uncertainty in the phase is determined by the Cram\'er-Rao bound
\cite{helstrom76}: 
\begin{equation}
 (\Delta\phi)^2 \geq \frac{1}{F(\phi)}\; ,
\end{equation}
where $F(\phi)$ is the Fisher information defined by
\begin{equation}
 F(\phi) = \sum_j \frac{1}{p(j|\phi)} \left[ \frac{\partial
     p(j|\phi)}{\partial \phi} \right]^2.
\end{equation}
When the input state is a perfect noon state, the Fisher information
is $F(\phi) = N^2$, and the Cram\'er-Rao bound yields $\Delta\phi \geq
1/N$. Up to a constant of proportionality, this bound is attained by
the measurement procedure outlined above.

\begin{figure}[t]
  \begin{center}
  \begin{psfrags} \psfrag{N}{\normalsize $N$} \psfrag{d}{\normalsize
$\Delta\phi$} \psfrag{a}{$1/\sqrt{N}$} \psfrag{b}{$1/N$}
\psfrag{c}{\large $\!\!\frac{e^{N\delta/2}}{N}$} \psfrag{x}{\footnotesize
0.05} \psfrag{0.1}{\footnotesize 0.1} \psfrag{0.15}{\footnotesize
0.15} \psfrag{0.2}{\footnotesize 0.2} \psfrag{0}{\footnotesize 0}
\psfrag{100}{\footnotesize 100} \psfrag{200}{\footnotesize 200}
\psfrag{300}{\footnotesize 300} \psfrag{400}{\footnotesize 400}
\epsfig{file=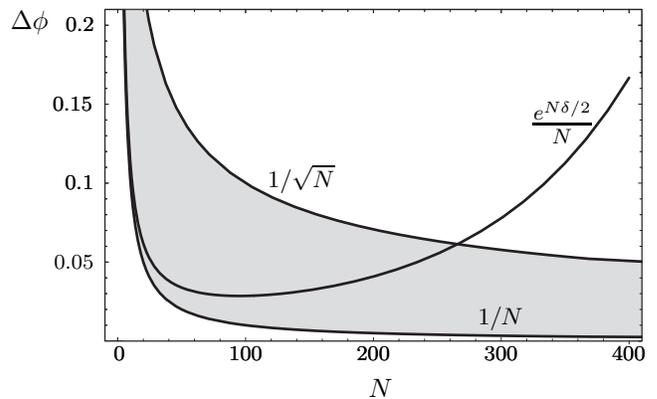,width=8.5cm}
  \end{psfrags}
  \end{center}
  \caption{Phase sensitivity $\Delta\phi$ of imperfect noon states
with noise parameter $\delta=0.02$ and $\phi=\pi/2N$ as a function of
$N$. The shaded area is bounded by  the standard quantum limit
$1/\sqrt{N}$ and the Heisenberg limit $1/N$. The minimum uncertainty
point is given by $N = 2/\delta$.}
  \label{fig:phase}
\end{figure}

When we take into account the Gaussian noise in the state preparation
process, the two conditional probabilities become
\begin{equation}
 p(\pm|\phi) = \frac{1}{2} \pm \frac{1}{2} \cos(N\phi)\, e^{-N\delta/2}.
\end{equation}
Consequently, the Fisher information is 
\begin{equation}
 F(\phi) = \frac{N^2 \sin^2(N\phi)}{e^{N\delta} - \cos(N\phi)}\, ,
\end{equation}
which is maximal when $\phi = \pi/2N$. The uncertainty in the phase at
this point is then:
\begin{equation}
 \Delta\phi \geq \frac{e^{N\delta/2}}{N}\, .
\end{equation}
This function exhibits a minimum at $N = 2/\delta$, as shown in
Fig.~\ref{fig:phase}. This means that the phase sensitivity $\delta$
of the optical control limits the size of the useful noisy noon states
that can be generated. As expected, when $\delta \rightarrow 0$ we
retrieve the Heisenberg limit. It should be mentioned that no
optimization of the phase estimation procedure has been performed.

\paragraph*{Bootstrapping}
If the number of photons in a useful noon states is limited by the
phase uncertainty as described above, then these states would be of
little use in metrology. However, we 
can use so-called {\em bootstrapping} to increase the effective noon
states to arbitrary photon number. The idea behind this technique is
to use (noisy) noon states to improve the phase uncertainty in the
optical control. For example, suppose that the phase shifters
producing the phases in Eq.~(\ref{eq:factors}) are implemented with 
delay lines, and the error $\Delta l$ in the delay is related to
$\sqrt{\delta}$ according to $\sqrt{\delta} = k \Delta l$, with $k$
the wave number. The
resulting noisy noon state can be used to re-evaluate the length of
the delay lines used in the state preparation process. If the error in
the length estimation using the noisy noon state is {\em smaller} than
the initial error $\sqrt{\delta}$, then the delay lines can be set
with a higher accuracy. Bootstrapping occurs when this higher accuracy
is used to tune smaller increments in the phase shifts and
consequently create a larger noisy noon state. This procedure can then be
repeated indefinitely. 

Clearly, for bootstrapping to work the minimum phase uncertainty
$\Delta\varphi_{\rm min}$ obtained by noon states must be smaller than
the phase uncertainty $\sqrt{\delta}$ in the apparatus: 
\begin{equation}\label{eq:boot} 
 \Delta\varphi_{\rm min} = \frac{e^{N\delta/2}}{N} < \sqrt{\delta} .
\end{equation} 
Since the minimum value of the phase uncertainty is reached when $N =
2/\delta$, we substitute this into Eq.~(\ref{eq:boot}) and solve the
inequality. We find that bootstrapping is possible when $\sqrt{\delta}
< 2/e$. Furthermore, if $\sqrt{\delta_0}$ is the initial phase
uncertainty  and $\sqrt{\delta_n}$ is the uncertainty in the $n^{th}$
iteration, the bootstrapping converges to zero super-exponentially: 
\begin{equation}
 \delta_n = \left( \frac{e}{2}\, \delta_0 \right)^{2^n}
 \quad\text{and}\qquad
 N_n = 2 \left( \frac{1}{e}\, N_0 \right)^{2^n}.
\end{equation}
For an initial phase uncertainty of 0.05~rad, the optimal noon state
contains ten photons. After two and three bootstrapping iterations,
the optimal noon state contains $\sim 180$ and $10^5$ photons, respectively. 

\paragraph*{Conclusion}
I have shown that the limits to optical phase control put a bound on
the size of the noon states that can be created with linear optics and
projective measurements, while still being able to perform
sub-shot-noise phase estimation. If the error in the phase control is
given by $\sqrt{\delta}$, then the maximum phase sensitivity in
standard Heisenberg-limited metrology is reached when $N =
2/\delta$. However, an adaptive bootstrapping technique can be used to
create high-fidelity noon states of arbitrary size (high-noon
states). Furthermore, this techique reduces the phase uncertainty
super-exponentially.  

\bigskip

This research is part of the QIP IRC www.qipirc.org (GR/S82176/01), and
was inspired by the LOQuIP workshop in Baton Rouge.

\end{document}